# Progress of Inorganic Scintillators for Future HEP Experiments

*Liyuan* Zhang*, and Ren-Yuan* Zhu[*]

California Institute of Technology,1200 E California Blvd, Pasadena, CA 91125, USA

**Abstract.** The Caltech HEP Crystal Lab has been actively investigating novel inorganic scintillators along the following three directions. Fast and radiation hard inorganic scintillators to face the challenge of severe radiation environment expected by future HEP experiments at hadron colliders, such as the HL-LHC and FCC-h. Ultrafast inorganic scintillators to face the challenge of unprecedented event rate expected by future HEP experiments searching for rare decays, such as Mu2e-II, and ultrafast time of flight system at hadron colliders. Cost-effective inorganic scintillators for the homogeneous hadron calorimeter concept to face the challenge of both electromagnetic and jet mass resolutions required by the proposed Higgs factory. We report novel materials along all directions: LuAG:Ce ceramic fibers for the HL-LHC, $Lu_2O_3$:Yb ceramic scintillators for ultrafast applications, and ABS:Ce and DSB:Ce glass scintillators for the proposed Higgs factory. The result of this investigation may also benefit nuclear physics experiments, GHz hard X-ray imaging, medical imaging, and homeland security applications.

## 1 Introduction

Total absorption electromagnetic calorimeters (ECAL) made of inorganic crystals provide the best energy resolution and detection efficiency for photons and electrons, so are the choice for those HEP experiments requiring the ultimate energy resolution. Novel crystal detectors are being discovered from academic research and by industry. They provide an important opportunity for future HEP detectors.

Following the priority research directions documented in the DOE basic research needs for HEP instrumentation [1], we have been actively investigating novel inorganic scintillators along the following three directions: 1) fast and radiation hard inorganic scintillators to face the challenge of severe radiation environment expected by future HEP experiments at hadron colliders, such as the HL-LHC and FCC-hh, where radiation damage is induced by ionization dose, protons and neutrons; 2) ultrafast inorganic scintillators to face the challenge of unprecedented event rate expected by future HEP experiments searching for rare decays, such as Mu2e-II, and ultrafast time of flight (TOF) system at future hadron colliders, and 3) cost effective inorganic scintillators for the homogeneous hadron calorimeter (HHCAL) concept to face the challenge of both electromagnetic and jet mass resolutions required by the proposed Higgs factory.

We report recent progress in all three directions and their potential applications for on-going calorimeter R&D. Radiation hard LuAG:Ce ceramic fibers and ultrafast $Lu_2O_3$:Yb ceramic scintillators for RADiCAL [2], and cerium-doped aluminoborosilicate (ABS:Ce) and BaO•2$SiO_2$ glass (DSB:Ce) glass scintillators for CalVision and the HHCAL concept [3].

Fig.1 shows the inorganic scintillator samples reported in this paper. They are 1) nine ϕ1 mm LuAG:Ce ceramic fibers produced by Shanghai Institute of Ceramics (SIC), including four 40 mm long, three 60 mm long, and two 120 mm long, 2) four $Lu_2O_3$:Yb ceramic scintillator samples produced by Radiation Monitoring Devices Inc. (RMD), including two 27×27×5 mm$^3$ (SN5280 and SN5281) and two 25 mm cubes (SLO-187 and LO-195); 3) three ABS:Ce glass samples provided by Institute of High Energy Physics (IHEP), including a 5 mm cube (Z-S), a 24 mm cube (Z-M) and a 25×25×60 mm$^3$ block (Z-L); and 4) a DSB:Ce glass block of 20×20×150 mm$^3$ provided by the 2nd Physics institute of Justus-Liebig University Giessen, Germany.

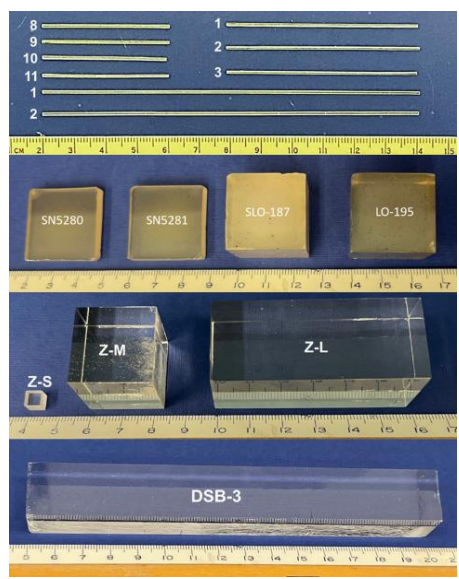

**Fig. 1**. Novel inorganic scintillators reported in this paper.

[*] Corresponding author: zhu@caltech.edu

We measured their photon-exited and X-ray-excited luminescence (PL and XEL), transmittance, pulse height spectra (PHS), γ-ray and α particle excited scintillation pulse shape, light output (LO) as a function of integration time, decay time (τ), and longitudinal light response uniformity (LRU) for long samples, as well as their degradation after γ-ray and proton irradiation.

## 2 LuAG:Ce ceramic fibers

Previous studies found that LuAG:Ce ceramic scintillators had a factor of two better radiation hardness than LYSO:Ce crystals against neutrons and protons [4]. We notice that the scintillation emission of LYSO:Ce matches well with the LuAG:Ce excitation. Fig. 2 shows one novel RADiCAL shashlik ECAL concept with LuAG:Ce fibers as wavelength shifter (WLS) to read LYSO:Ce scintillation light. Combined with W absorber and SiPM-based readout, this configuration provides an ultracompact, radiation-hard ECAL concept with a longitudinal segmentation capability.

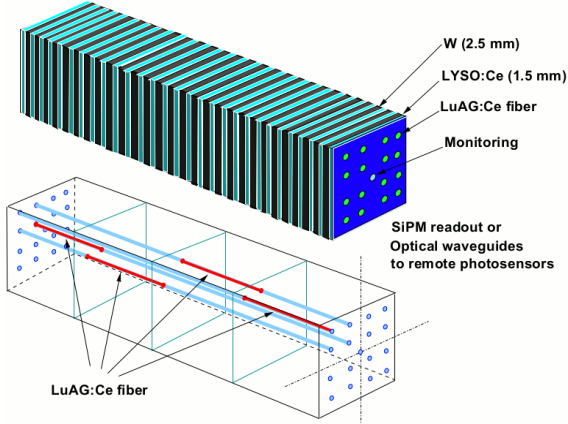

**Fig. 2.** One RADiCAL concept consisting of LYSO:Ce and W plates and LuAG:Ce WLS fibers.

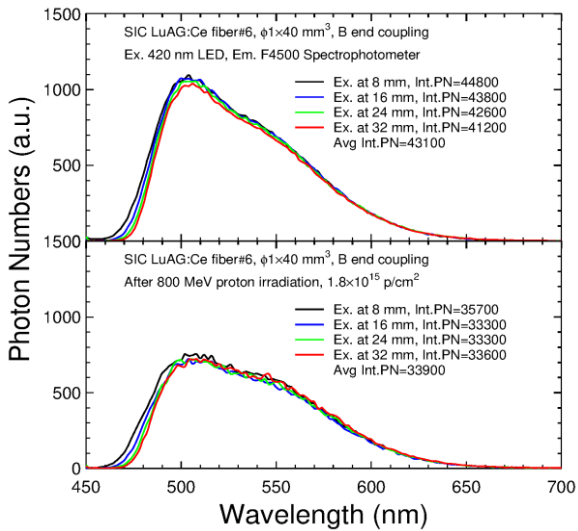

**Fig. 3.** PL spectra of the φ1×40 mm$^3$ LuAG:Ce ceramic fiber #6 measured before (top) and after (bottom) $1.8\times10^{15}$ p/cm$^2$.

Three φ1×40 mm$^3$ LuAG:Ce fibers, #10, #9 and #6, were irradiated by 800 MeV protons in the experiment LANSCE-9168 at the blue room of LANSCE with fluence of 0.19, 3.0 and $1.8\times10^{15}$ p/cm$^2$, respectively. Fig. 3 shows the PL spectra measured with 420 nm LED excitation at 8, 16, 24 and 32 mm from the end coupling to photodetector before (top) and after (bottom) $1.8\times10^{15}$ p/cm$^2$ for the φ1×40 mm$^3$ LuAG:Ce ceramic sample #6. Fig. 4 shows the corresponding integrated PL intensity as a function of the LED excitation position. The average PL intensity decreases by 19%, 21% and 28% for the samples #10, #9 and #6 respectively, showing excellent radiation hardness against protons after $1.8\times10^{15}$ p/cm$^2$. This result is consistent with our previous measurements for LuAG:Ce ceramics [4].

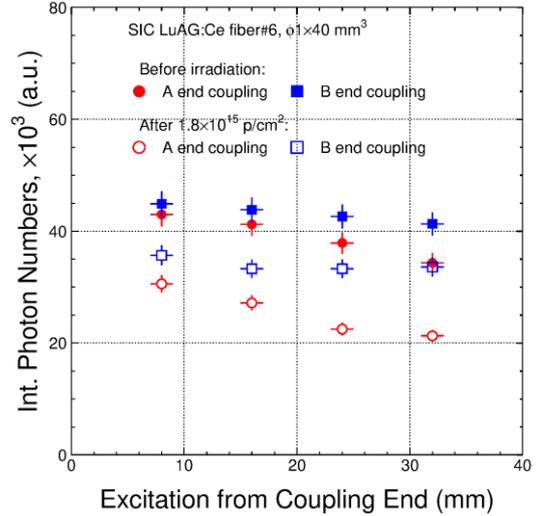

**Fig. 4.** Integrated PL intensity is shown as a function of the excitation position for the φ1×40 mm$^3$ LuAG:Ce ceramic fiber #6 with alternative end coupling to photodetector.

## 3 Lu$_2$O$_3$:Yb ceramic scintillators

Inorganic scintillators activated by charge transfer luminescence Yb$^{3+}$ are considered promising ultrafast scintillator to break the ps timing barrier for future HEP TOF and calorimetry application.

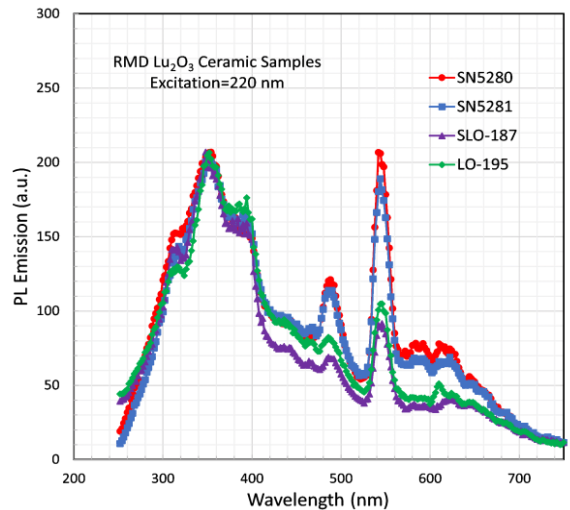

**Fig. 5.** PL spectra of four Lu$_2$O$_3$:Yb ceramic samples.

Inorganic scintillators in ceramic form are also potentially more cost-effective than crystals because of their lower fabrication temperature and no need for aftergrowth mechanical processing. With high density (9.4 g/cm$^3$) and ultrafast decay (< 1 ns) Lu$_2$O$_3$:Yb ceramic scintillators may find applications for ultrafast TOF and calorimetry. Fig. 5 shows photoluminescence for four Lu$_2$O$_3$:Yb samples. All samples have a consistent broad charge-transfer emission band at 300-400 nm, and narrow emission peaks at longer than 400 nm, which are attributed to rare earth elements in raw materials. Fig. 6 shows PHS measured for the sample SN5280 with Na-22 excitation and a coincidence trigger for three integration time of 45, 200 and 1,000 ns.

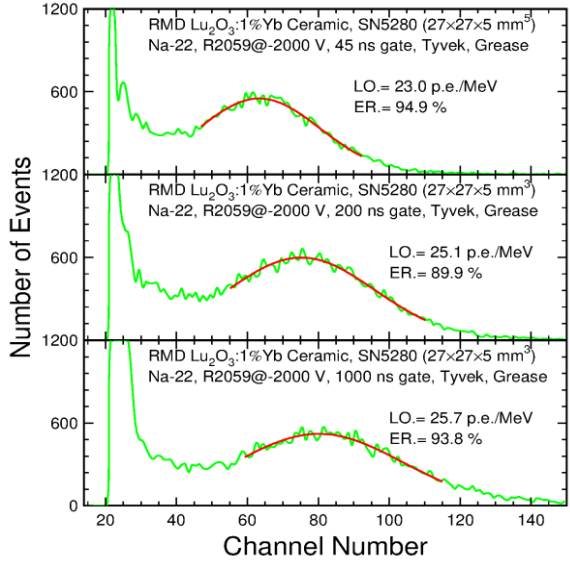

**Fig. 6.** PHS of the Lu$_2$O$_3$:Yb ceramic sample SN5280 measured with three integration time: 45, 200 and 1,000 ns.

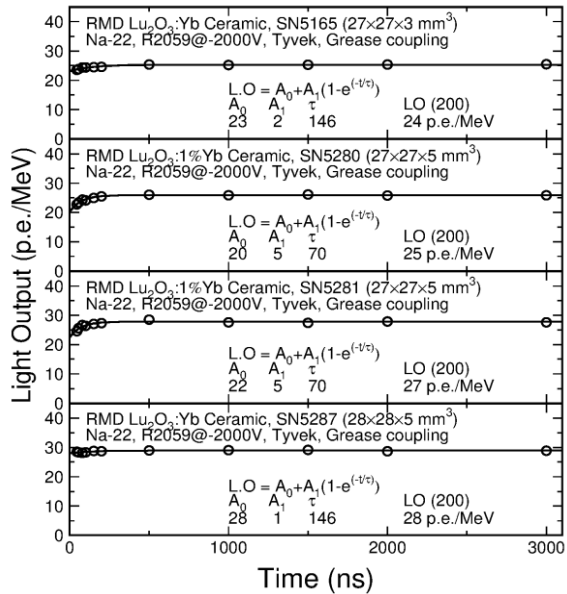

**Fig. 7.** LO is shown as a function of integration time for four Lu$_2$O$_3$:Yb samples with exponential fit to extract an ultrafast scintillation component (A$_0$) and a slow component (A$_1$).

Fig. 7 shows LO as a function of integration time and corresponding exponential fit to extract an ultrafast component (A0) and a slow component (A1) for four Lu$_2$O$_3$:Yb ceramic samples. While all samples show some level of slow component, the sample SN5287 shows the best fast/slow ratio (A$_0$/A$_1$). Its high light yield in the first nanosecond and high fast/slow ratio provides an effective sensor for future TOF detector for future high-rate HEP experiments.

## 4 ABS and DSB glass scintillators

Bright and UV-transparent heavy inorganic scintillators are required for the HHCAL concept [4]. Because of the huge volume required, the key issue is cost-effective inorganic scintillator. With density of 6 and 4.3 g/cm$^3$ and expected mass-production cost of about $1/cc, the ABS:Ce [5] and DSB:Ce [6] glass scintillators are promising for the HHCAL concept.

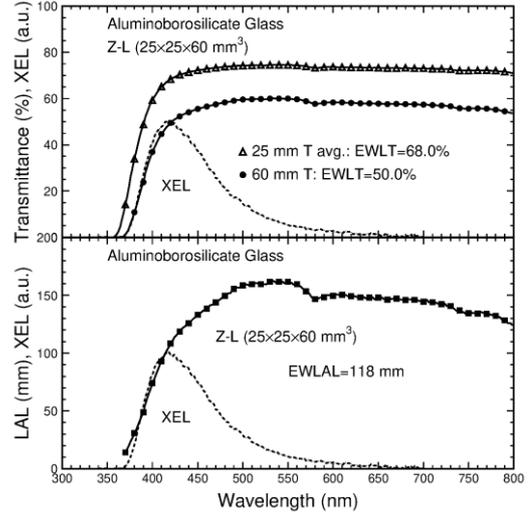

**Fig. 8.** Transmittance (top), XEL and LAL (bottom) spectra are shown for the 25×25×60 mm$^3$ ABS:Ce glass sample Z-L.

The top plot of Fig. 8 shows the longitudinal and transverse transmittance spectra for the 25×25×60 mm$^3$ ABS:Ce sample Z-L. Also shown in the figure is its XEL emission spectrum and the numerical values of its emission-weighed longitudinal transmittance (EWLT) measured for different optical pathlength. The bottom plot of Fig. 8 shows the light attenuation length (LAL) as a function of wavelength calculated from the difference between the transverse and longitudinal transmittance, and the numerical value of the emission weighted LAL (EWLAL) of 11.8 cm.

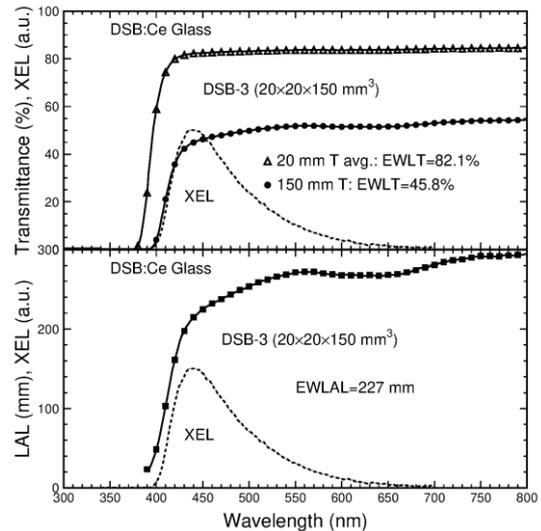

**Fig. 9.** Transmittance (top), XEL and LAL (bottom) spectra are shown for the 20×20×150 mm$^3$ DSB-3 glass sample.

Fig. 9 shows the corresponding plots for a 20×20×150 mm$^3$ DSB-3 sample with EWLAL of 22.7 cm, indicating that its optical quality is better than ABS:Ce sample Z-L. Their much shorter LAL as compared to a few meters of typical crystal scintillators indicate that the optical quality of current glass scintillators needs to be further improved for the total absorption calorimeter application.

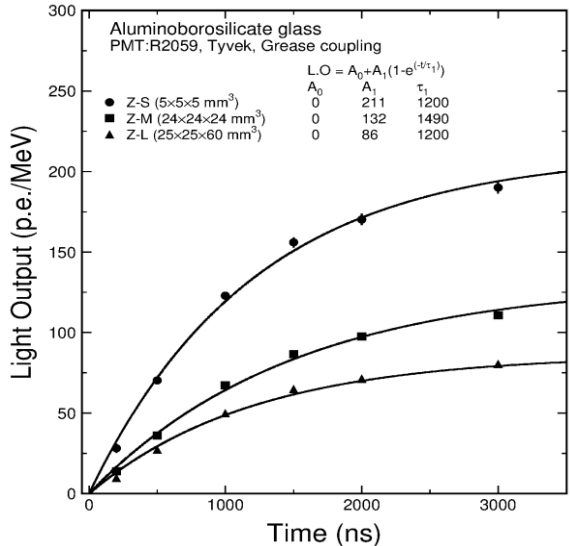

**Fig. 10.** LO as a function of integration time is shown for three ABS glass samples.

Fig. 10 shows LO as a function of integration time for three ABS samples with consistent decay time of 1.2~1.5 μs.

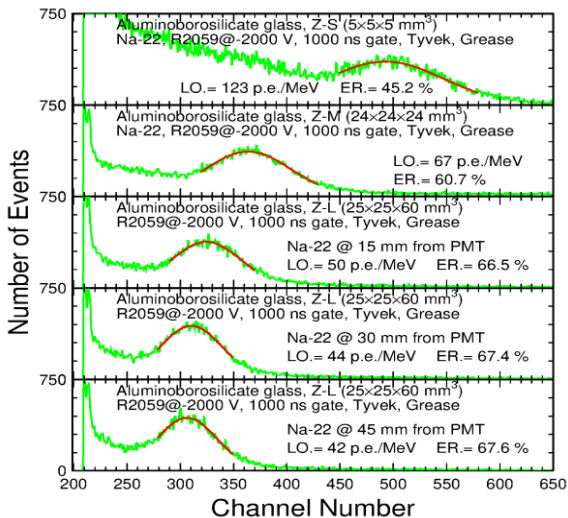

**Fig. 11.** PHS measured with one μs gate is shown for the three ABS glass samples.

Fig. 11 shows PHS for the three ABS samples measured with one μs gate. Because of internal absorption the LO value decreases significantly from 210 p.e./MeV for Z-S to 86 p.e./MeV for Z-L. The bottom three PHS plots are for the 60 mm Z-L ABS samples with Na-22 source excitation along its longitudinal axis at 15, 30 and 45 mm from the photodetector with an average of 45 p.e./MeV and a rms value of 7.4% representing its light response non-uniformity. The corresponding values are 500 ns, 80 p.e./MeV and 10.3% for the 20×20×150 mm$^3$ DSB:Ce glass sample.

## 5 Summary

We report recent progress in novel inorganic scintillators. Bright, fast and radiation hard LuAG:Ce ceramic fibers are under development for the HL-LHC and FCC-hh. Lu$_2$O$_3$:Yb ceramic scintillators are under development for ultrafast TOF and calorimetry applications. ABS:Ce and DSB:Ce glass scintillators are under development for the proposed Higgs factory. The result of this investigation may also benefit nuclear physics experiments, GHz hard X-ray imaging, medical imaging, and homeland security applications.

## Acknowledgements

Doctors A. Wu, J. Li and L. Su of SIC, L.S. Pandian, Y. Wang and J. Glodo of RMD, Q. Sen of IHEP, V. Dormenev and R. Novotney of Giessen university provided samples investigated in this paper. This work was supported in part by the U.S. Department of Energy under Grant DE-SC0011925 and DE-SC0024094.